\documentstyle[preprint,aps]{revtex}
\tighten
\draft
\begin{document}
        
\title{\bf Low-lying $2^+$ states in neutron-rich oxygen isotopes in 
quasiparticle random phase approximation}  

\author{\rm E. Khan and Nguyen Van Giai}

\address{ Institut de Physique Nucl\'eaire, IN$_{2}$P$_{3}$-CNRS,  
91406 Orsay Cedex, France\\ }

\maketitle

\begin{abstract}
The properties of the low-lying, collective
2$_1^+$ states in neutron-rich oxygen isotopes are investigated in the
framework of
self-consistent microscopic models with effective Skyrme interactions.
In RPA the excitation
energies $E_{2^+_1}$ can be well described but the transition probabilities
are much too small as compared to experiment. Pairing correlations are then
accounted for by performing quasiparticle RPA calculations. This
improves considerably the predictions of B(E2)
values and it enables one to calculate more reliably the ratios $M_n/M_p$ of
neutron-to-proton transition amplitudes. A satisfactory agreement with the
existing experimental values of $M_n/M_p$ is obtained. 
\end{abstract}


\newpage
The prospects of nuclear physics studies with nuclei far from stability open 
up a wide range of possibilities for refining our understanding of nuclear 
properties in terms of microscopic descriptions and effective nucleon-nucleon 
interactions. One of the important aspects is the ability to disentangle 
neutron and proton contributions to collective transitions between low-lying 
states and the ground state. Experimentally, this can be done in a
phenomenological way \cite{Ber83} by combining the 
information obtained in measurements involving various hadronic probes and 
purely electromagnetic probes. For instance, 
numerous experimental studies have been
   performed on $^{18}$O using different hadronic probes like
   proton\cite{esc74}, neutron\cite{grab80}, 
or pion\cite{iver79,see88} scattering.    
   More recently, proton scattering on $^{20}$O\cite{jew98} yielded
   information on the first $2^+$ state in this 
unstable neutron-rich oxygen isotope. On the
   other hand, studies involving only electromagnetic properties such as 
electron scattering\cite{kel86},  
Coulomb excitation or lifetime
   measurement\cite{ram87} have also been done for these nuclei. While the 
excitation processes of purely electromagnetic nature are sensitive 
only to the protons 
and give access to the proton transition amplitudes and transition densities, 
the hadronic processes are sensitive to both proton and neutron transition 
densities. Therefore, it is possible by a combined analysis of the data from 
electromagnetic and hadronic processes to determine experimentally for a 
given excited state the transition amplitudes $M_p$ and $M_n$ corresponding 
to protons and neutrons, respectively\cite{Ber83}. Proton scattering
experiments yielding $M_n/M_p$ values have been recently performed on
neutron-rich sulfur and oxygen isotopes \cite{jew98,Kel97,Mar99}.   

In this work, we present microscopic calculations of low-lying $2^+$ states in 
neutron-rich oxygen isotopes. These calculations are based on effective Skyrme 
interactions and they are performed in the framework of the random phase 
approximation (RPA) and the quasiparticle random phase approximation (QRPA). In
microscopic models the properties of the states depend on two main inputs, the 
single-particle spectrum and the residual two-body interaction. In the present 
approach these two features are linked since the same effective interaction 
determines the Hartree-Fock (HF) single-particle spectrum and the residual 
particle-hole interaction. This approach has proved to be an efficient mean 
to predict properties of collective excitations like giant resonances\cite{ngu81} 
and it has also been used for calculating low-lying collective states in 
closed-shell nuclei\cite{liubrown76}. 

In unstable nuclei we usually don't deal 
with closed-shell or closed-subshell systems and therefore, the HF and RPA 
calculations must be done with additional approximations. The HF calculations 
are carried out assuming spherical symmetry and using the standard filling 
approximation with equal occupation numbers for all $(jm)$-substates of the 
partially filled $j$-subshell. The RPA calculations are then be carried out 
taking into account these occupation numbers. However, pairing correlations 
can be important in such nuclei and they must be taken into account. Their 
effects will be described by HFBCS calculations for the ground states and by 
QRPA calculations for the excitation spectra.  

The HF and HFBCS method in spherical nuclei with Skyrme interactions is 
well-known\cite{vau73,bei74}. For the pairing interaction we simply choose 
a constant gap given by\cite{boh69}:
\begin{equation}\label{eq:de}
  	\Delta = 12. A^{-\frac {1}{2}} {\rm MeV}~.
\end{equation}
In a more realistic treatment of the pairing, the gap would depend on the
single-particle state considered and it would tend to zero when the subshell
is far from the Fermi level. Thus, in the constant gap approximation it is
necessary to introduce a cut-off in the single-particle space. 
Above this cutoff subshells don't
participate to the pairing effect. In the case of oxygen isotopes,
we choose the BCS subspace to include the $1s, 1p$ and $2s-1d$ major shells. 

The results of the HF and HFBCS calculations performed for the nuclei 
$^{18,20,22}$O using typical Skyrme interactions are summarized in 
Table \ref{bcs}. 
For all the interactions, the binding energies per particle decrease with 
increasing neutron number but SGII has a different behavior as compared to the 
other interactions and it predicts larger values of $B/A$. The pairing 
correlations decrease $B/A$ by about 4\% in all cases. The neutron radii 
increase substantially from $^{18}$O to $^{22}$O and the 
proton radii also increase slightly as a result of neutron-proton 
attraction. The effect of neutron pairing correlations is to redistribute 
neutron densities to the tail region and therefore, this leads to a small 
increase in $r_n$ (and also in $r_p$ for the reason mentioned above). 

The HF-RPA model with Skyrme effective forces is also well-known. We only 
mention that in this work we solve the RPA equations in configuration 
space, choosing the particle-hole space so as to exhaust the energy-weighted 
sum rules. The continuous part of the single-particle spectrum is discretized 
by diagonalizing the HF hamiltonian on a harmonic oscillator basis\cite{colo}.  
To generalize the HF-RPA to the QRPA model we follow the standard 
procedure\cite{rowe}. Denoting by $a^{\dagger}_{\alpha}$, $a_{\alpha}$ the 
creation and annihilation operators of a particle in a HF state 
$\alpha = (j_\alpha, m_\alpha)$ and 
by $c^{\dagger}_{\alpha}$, $c_{\alpha}$ the corresponding operators for a 
quasiparticle state, we have:  
\begin{equation}\label{eq:hf}
 c^{\dagger}_{j_{\alpha}m_{\alpha}} = u_{\alpha} a^{\dagger}
_{j_{\alpha}m_{\alpha}} -
 v_{\alpha} (-1)^{j_{\alpha} + m_{\alpha}}a_{j_{\alpha}-m_{\alpha}}~,
\end{equation}
 where the BCS amplitudes $u_\alpha, v_\alpha$ satisfy the normalization 
condition: 
\begin{equation}\label{eq:no}
   u_{\alpha}^2 + v_{\alpha}^2 = 1~.
 \end{equation}
These amplitudes are determined by solving the BCS equations. One can then build 
the two-quasiparticle creation operators in an angular momentum coupled 
scheme:
\begin{equation}\label{eq:op}
  C^\dagger_{\alpha\beta}(JM) = (1+\delta_{\alpha\beta})^{-1/2}
  \sum_{m_\alpha m_\beta} (j_\alpha j_\beta m_\alpha
  m_\beta|JM) c^{\dagger}_{\alpha}c^{\dagger}_{\beta}~.
\end{equation}
In QRPA the nuclear excitations correspond to phonon operators which are linear 
combinations of two-quasiparticle creation and annihilation operators: 
\begin{equation}\label{eq:ex}
  Q^{\nu\dagger}(JM) = \sum_{\alpha\geq\beta}
  X^\nu_{\alpha\beta}(J)C^\dagger_{\alpha\beta}(JM)
  + (-1)^M  Y^\nu_{\alpha\beta}(J)C_{\alpha\beta}(J-M)~.
\end{equation}
Making use of the condition that the QRPA ground state $\vert \tilde 0 \rangle$ 
is a vacuum of phonon:
\begin{equation}\label{eq:vac}
Q^{\nu}(JM)\vert \tilde 0 \rangle = 0~,
\end{equation}
one can then derive the QRPA equations whose solutions yield the excitation 
energies $E_\nu$ and amplitudes $X^\nu_{\alpha\beta}, Y^\nu_{\alpha\beta}$ of 
the excited states.

An important quantity that characterizes a given state $\nu = (E_\nu, LJ)$ is its 
transition density:
\begin{equation}\label{eq:tran}
\delta\rho^\nu({\bf r}) \equiv  \langle \nu \vert \sum_i \delta 
({\bf r - r_i})\vert \tilde 0 \rangle ~,
\end{equation}
and a similar definition of the neutron (proton) transition density 
$\delta\rho_n^\nu$ ($\delta\rho_p^\nu$) with the summation in eq.(\ref{eq:tran}) 
restricted to neutrons (protons). In QRPA the radial part of the transition 
density is:
\begin{equation}\label{eq:tr}
\delta\rho^{\nu}(r)=\sum_{\alpha\geq\beta} \varphi_{\alpha}(r)\varphi^{*}_{\beta}(r)
<\beta||Y_{L0}||\alpha> \lbrace{X^\nu_{\alpha\beta}(J)-
Y^\nu_{\alpha\beta}(J)}\rbrace \lbrace{u_{\alpha}v_{\beta}+(-1)^Jv_{\alpha}
u_{\beta}}\rbrace~,
 \end{equation}
where $\varphi_{\alpha}(r)$ is the radial part of the wavefunction of the
quasiparticle state $\alpha$. 
As an example the QRPA 
neutron and proton transition densities of the first $2^+$ state in $^{20}$O 
calculated with the interaction SGII are shown in Fig.1. 
The neutron transition density is shifted outwards as 
compared to the proton transition density due to 
the presence of a neutron skin. Clearly, the two transition densities do not scale 
like $N/Z$ as it is sometimes assumed and this has quantitative consequences as 
we shall see below. 

The neutron and proton matrix elements $M = <\nu|r^L Y_{L0}|\tilde0>$ of a multipole operator  
are obtained by integrating
the corresponding transition densities over r : 
\begin{equation}\label{eq:mn}
M_{n,p} = \int \delta\rho^{\nu}_{n,p}(r) r^{L+2}dr~,
\end{equation}
and the reduced electric multipole transition probabilities are calculated as
\begin{equation}\label{eq:be}
 B(EL)_{n,p}  = |M_{n,p}|^2~.	
\end{equation}

We have calculated the $J^\pi$ = 2$^+$ states in $^{18,20,22}$O using RPA and QRPA 
with SIII, SGII and SLy4 interactions. The energies and $B(E2)_p$ values 
of the first 2$^+$ states are shown 
in Fig.2 together with the existing experimental values in $^{18}$O 
and $^{20}$O. The data come from experiments involving 
 electromagnetic processes such as Coulomb excitation or lifetime
 measurement\cite{ram87}. For the $E_{2^{+}}$ energies, 
standard RPA reproduces very well experimental values
especially for SGII and SLy4. On the other hand, QRPA deteriorates this 
agreement and it predicts the first 2$^+$ states at somewhat higher energies. 
The theoretical prediction of the energies of low-lying states in models 
based on a HF or HFBCS mean field is a delicate task because these energies 
are sensitive to the spin-orbit part of the mean field while the spin-orbit 
component of the two-body effective interaction is not so well determined. 
For the three interactions used here there is a clear difference in the QRPA 
$E_{2^{+}}$ energies between SGII and the other interactions. 
In the case of B(E$_2$)$_p$  values RPA predicts too small values with the three
interactions. However, the agreement becomes very good for QRPA both in $^{18}$O 
and $^{20}$O, especially for SGII and SLy4 interactions. 
This indicates that pairing effects are 
important for transition probabilities of the first 2$^+$ state in these nuclei. 

In Fig.3 are shown the ratios $M_n/M_p$ calculated with the three interactions 
within RPA and QRPA. On the same figure are displayed the experimental 
values taken
 from ref \cite{jew98}. It can be seen that the RPA results are somewhat larger 
than those of QRPA due to the very small B(E$_2$)$_p$ values obtained in RPA. 
In comparison with the data, the QRPA results are in good agreement in $^{20}$O 
whereas they are slightly too large in $^{18}$O. The three interactions predict 
different $A$ dependence in these oxygen isotopes for the QRPA ratios. 
It would be interesting to obtain the experimental value of $M_n/M_p$ in $^{22}$O 
as well as the corresponding B(E2) transition probability. 
If one assumes the quadrupole excitation to be purely isoscalar the ratio 
$M_n/M_p$ would be equal to $N/Z$. Taking as a guideline the QRPA results 
calculated with SGII one sees that the ratio of neutron-to-proton transition 
amplitudes is about (1.8 - 2.0)$N/Z$, thus indicating that the low-lying 
2$^+$ state has an important isovector component. This is not surprising since 
in these neutron-rich nuclei there are neutron particle-hole configurations at 
low energy which have no counterpart on the proton side and therefore, these 
neutron configurations necessarily introduce both isoscalar and isovector 
type of excitations. 

In summary, we have investigated the properties of the low-lying, collective
2$_1^+$ states in neutron-rich oxygen isotopes in the framework of
self-consistent microscopic models. Within the RPA model the excitation
energy $E_{2^+_1}$ can be well described but the transition probabilities
are much too small as compared to experiment. In these open subshell nuclei
the pairing correlations can be important and therefore, we have extended
the previous model to the HFBCS approximation at the mean field level and we
have performed QRPA calculations for the excited  states. The quasiparticle
microscopic description improves considerably the predictions of B(E2)
values and it enables us to calculate more reliably the ratios $M_n/M_p$ of
neutron-to-proton transition amplitudes. These calculated values
differ noticeably from the naive $N/Z$ estimate and they are in
satisfactory agreement with experiment. However, the QRPA overestimates the
$E_{2^+_1}$ energies. The sensitivity of positive parity low-lying states to
the effective interaction should give one a handle on some specific
components of the force, for instance the two body spin-orbit part. 

Further proton scattering results on $^{18}$O and $^{20}$O will be available
soon\cite{kha99}, yielding energies and $M_{n}/M_{p}$ ratio for the first
2$^+$ and 3$^-$ states. 
It would also be useful to perform experiments on more neutron-rich oxygen
isotopes to
establish firmly the trend of $M_{n}/M_{p}$ as a function of $N/Z$. 

\vspace{0.5cm}
\noindent
We would like to thank G. Col\`o, T. Suomij\"arvi and C. Volpe for useful
discussions.

\newpage\eject
\noindent
{\Large \bf Figure captions}
\vskip 48pt
\par\noindent

{\bf Figure 1.}
Neutron and proton transition densities of the first 2$^+$ state in
$^{20}$O, calculated in QRPA with interaction SGII.

{\bf Figure 2.}
Energies and B(E2)$_p$ values of the first 2$^+$ states in oxygen isotopes.
Open and black symbols correspond to RPA and QRPA calculations,
respectively. Three effective interactions are used: SIII\cite{bei74}
(circles), SGII\cite{ngu81} (triangles), SLy4\cite{chab98} (stars).
Experimental values are shown as crosses, with error bars for B(E2)$_p$. 

{\bf Figure 3.}
The $M_n/M_p$ ratios in oxygen isotopes. The notations are the same as in
Fig. 2.
\begin{table}[h]
\begin{tabular}{|c|c||c|c|c||c|c|c||c|c|c||}
\hline
 & & \multicolumn{3}{c||}{$^{18}$O} & \multicolumn{3}{c||}{$^{20}$O} &
 \multicolumn{3}{c||}{$^{22}$O}
\\ \cline{3-11}
 & &  r$_n$ (fm)& r$_p$ (fm)  & B/A (MeV)
 &  r$_n$ (fm)& r$_p$ (fm)  & B/A (MeV) 
 &  r$_n$ (fm)& r$_p$ (fm)  & B/A (MeV) 
 \\ \hline
SIII &HF& 2.77 & 2.65  & 7.74& 2.89 & 2.67  & 7.58 & 2.97 & 2.69 & 7.48
\\ \cline{2-11}
     &BCS& 2.79 & 2.67  & 7.43& 2.91 & 2.69  & 7.28 & 3.03 & 2.71 & 7.16
\\ \hline
SGII &HF& 2.76 & 2.64  & 8.30& 2.87 & 2.65  & 8.15 & 2.95 & 2.66 & 8.07
\\ \cline{2-11}
    &BCS& 2.80 & 2.66  & 7.95& 2.92 & 2.68  & 7.79 & 3.02 & 2.69 & 7.69   
\\ \hline
SLy4 &HF& 2.83 & 2.69  & 7.74& 2.94 & 2.70  & 7.57 & 3.02 & 2.71 & 7.46
\\ \cline{2-11}
     &BCS& 2.85 & 2.72  & 7.42& 2.98 & 2.73  & 7.25 & 3.08 & 2.74 & 7.13
\\ \hline
\end{tabular}
\caption{\label{bcs}}
Neutron and proton r.m.s. radii, and binding energy per
particle in the nuclei $^{18}$O, $^{20}$O and $^{22}$O 
calculated with interactions SIII\cite{bei74}, SGII\cite{ngu81} and 
SLy4\cite{chab98}. The rows labeled HF and BCS show Hartree-Fock and 
Hartree-Fock-BCS results, respectively. 
\end{table}


\begin{references}
\bibitem{Ber83} A.M. Bernstein, V.R. Brown and V.A. Madsen, 
{\it Comments Nucl. Part. Phys.} {\bf 11} 
(1983) 203.
\bibitem{esc74} J.L. Escudi\'e et al., {\it Phys. Rev. }{\bf C10} (1974) 1645.
\bibitem{grab80} P. Grabmayr, J. Rapaport and R.W. Finlay, 
{\it Nucl. Phys. }{\bf A350} (1980) 167.
\bibitem{iver79} S. Iversen et al., {\it Phys. Lett. }{\bf 82B} (1979) 51.
\bibitem{see88} S.J. Seestrom-Morris et al., {\it Phys. Rev. }{\bf C37} (1988) 2057.
\bibitem{jew98} J.K. Jewell et al., {\it Phys. Lett.} {\bf 454B} (1999) 181.
\bibitem{kel86} J. Kelly et al.,  {\it Phys. Lett. }{\bf 169B} (1986) 157.
\bibitem{ram87} S. Raman et al.,  {\it Atomic Data and Nuclear Data Tables }
36 (1987) 1.
\bibitem{Kel97} J.H. Kelley et al., {\it Phys. Rev. }{\bf C56} (1997) R1206.
\bibitem{Mar99} F. Mar\'echal et al., {\it to be published in Phys. Rev. }{\bf C} 
\bibitem{ngu81} Nguyen Van Giai and H. Sagawa, {\it Nucl. Phys.} {\bf A371} (1981) 1.
\bibitem{liubrown76} K.F. Liu and G.E. Brown, {\it Nucl. Phys.} {\bf A265} 
(1976) 385.
\bibitem{vau73} D. Vautherin,  {\it Phys. Rev. }{\bf C7} (1973) 6.
\bibitem{bei74} M. Beiner, H. Flocard, Nguyen Van Giai and P. Quentin, 
{\it Nucl. Phys. }{\bf A238} (1975) 29.
\bibitem{boh69} A. Bohr and B. Mottelson,  {\it Nuclear structure} (Benjamin,
New York, 1969) Vol. 1, p. 169.
\bibitem{chab98} E. Chabanat et al.,  {\it Nucl. Phys. }{\bf A635} (1998) 231.
\bibitem{colo} G. Col\`o, Nguyen Van Giai, P.F. Bortignon and R.A. Broglia,
{\it Phys. Rew.} {\bf C50} (1994) 1496.
\bibitem{rowe} D.J. Rowe, {\it Nuclear Collective Motion: Models and Theory}
(Methuen, 1970).
\bibitem{kha99} E. Khan et al., {\it Proceeding of the XXXVII international
winter meeting on nuclear physics} (1999) Bormio, Italy.
\end{references}
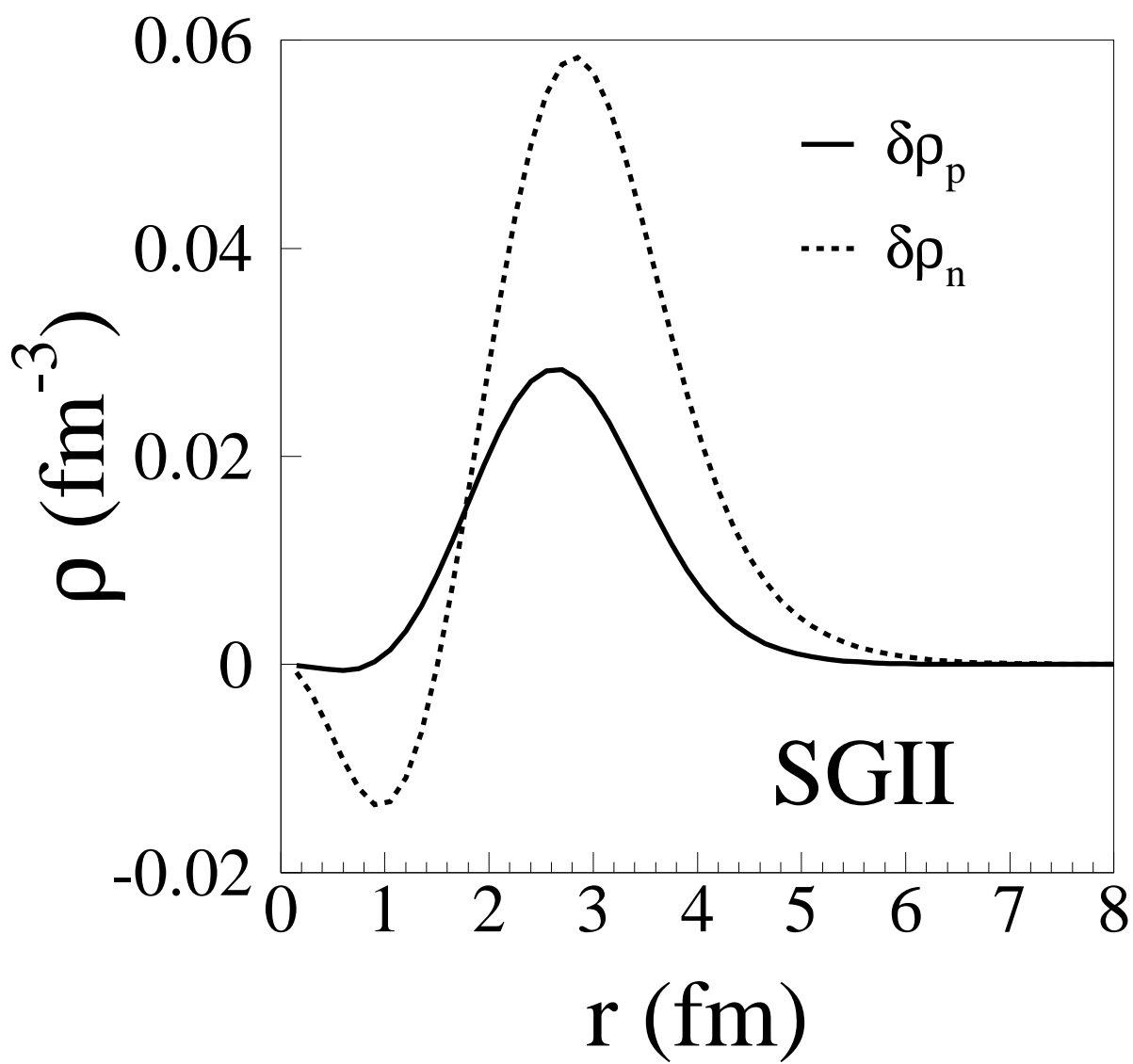
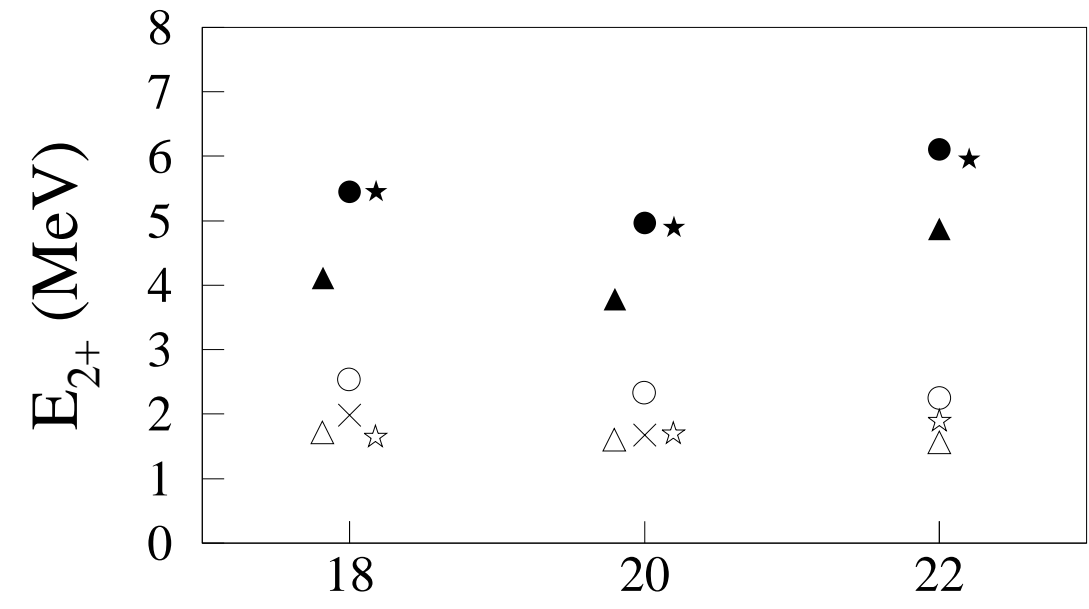
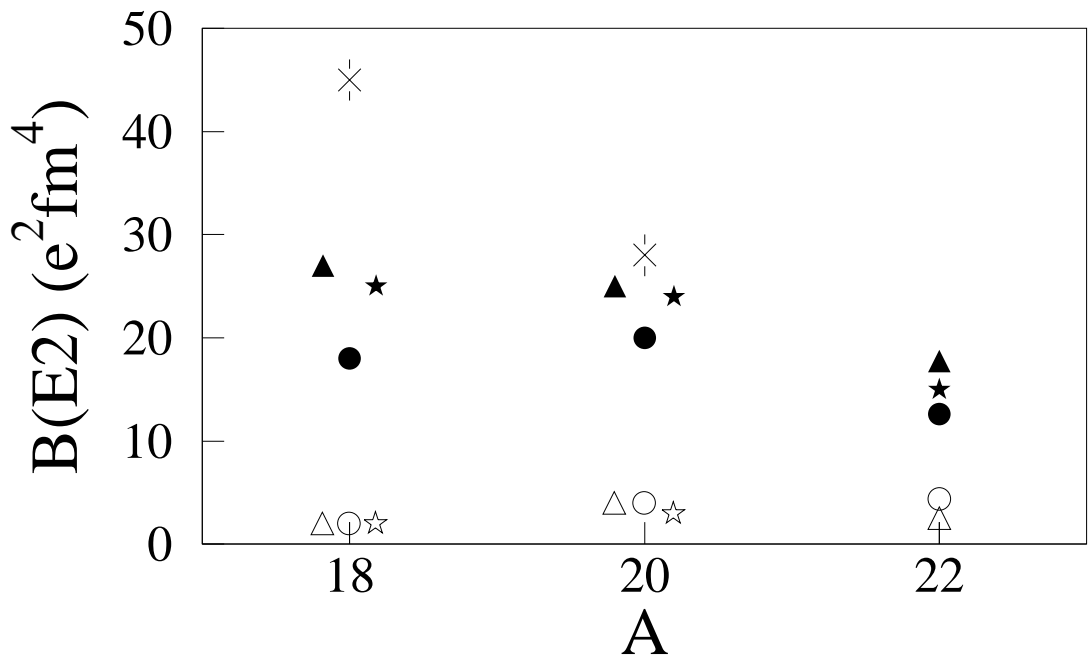
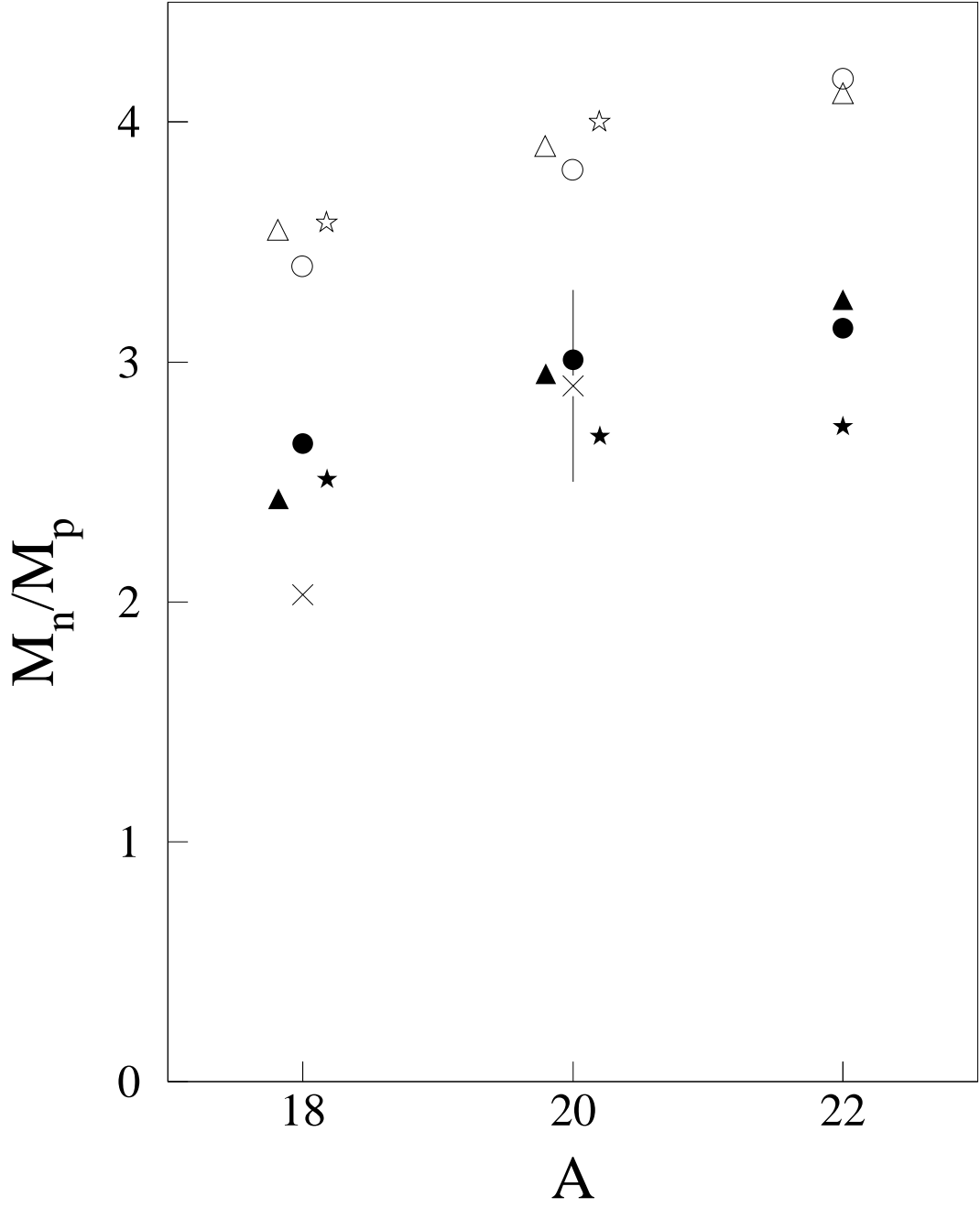
\end{document}